# DEFINITION OF A NEW LEVEL ONE TEST CASE MEASUREMENTS OF EQUILIBRIUM RADIATION FROM AN INDUCTIVELY COUPLED PLASMA IN THE NEAR-UV TO NEAR-IR SPECTRAL REGION FOR A TITAN-TYPE $N_2$-$CH_4$ MIXTURE. PRELIMINARY RESULTS

**Vacher D.(1), André P.(1), Lino da Silva M.(2), Dudeck M.(3), Faure G.(1)**

*(1)LAEPT, CNRS, Clermont-Ferrand, France, Damien.VACHER@laept.univ-bpclermont.fr*
*(2)Instituto de Plasmas e Fusao Nuclear, Instituto Superior Técnico, Lisboa, Portugal*
*(3)Institut Jean Le Rond d'Alembert, University of Paris 6, Paris, France*

## ABSTRACT

A test case, using a similar methodology and experimental set-up than previous test case TC5-Level 1 [1;2], is presented in this paper. An inductively coupled plasma torch, working at atmospheric pressure, is used to create $N_2$-$CH_4$ Titan-like plasma (98%$N_2$ - 2%$CH_4$). The operating frequency and power are 64 MHz and 3 kW respectively. This kind of apparatus allows obtaining plasma in chemical and quasi-thermal equilibrium The spectral measurements cover the [300-800] nm range and are performed inside the induction coil. Each interesting spectrum is calibrated and compared to the line-by-line spectral code SPARTAN used for the simulation of the radiative emission of entry-type plasma.

Finally, a discussion is proposed about the nucleation phenomenon which is occurred in the ICP torch with the $N_2$-$CH_4$ plasma. Preliminary studies show the synthesis of nanostructured carbon on the quartz tube.

## 1. INTRODUCTION

Different levels of test-cases have been defined in the frame of the ESA-CNES radiation working group. Such test-cases are useful for the evaluation and validation of radiation models used for atmospheric entry applications.

Test case n° 1 is devoted to the validation of models of line-by-line spectroscopic codes and of spectral data used for line-by-line radiation calculations. As the simplest case is the modelling of plasma in thermo-chemical equilibrium, an ICP torch, which allows obtaining such conditions, is used. Test case 1 was initially intended to validate measurements in air plasma. This new level of test case 1 proposes a spectroscopic study of plasma formed with a Titan-like atmosphere gas mixture, which will be useful for the validation of spectral databases (SPARTAN code).

The outline of the paper is as follows. In section 2, all the characteristics of the experimental set-up are given. In section 3, after a discussion concerning the sustained conditions of the plasma when methane is injected into a pure nitrogen plasma, experimental spectra in the range [320-840] nm are presented and an estimation of the axial temperature is given. In section 4, a discussion is proposed about the nucleation phenomenon which is occurred in the ICP torch with the $N_2$-$CH_4$ plasma. Preliminary studies show the synthesis of nanostructured carbon on the inner surface of the quartz tube. The chemical equilibrium composition of a $N_2$-$CH_4$ Titan-like plasma (98%$N_2$ - 2%$CH_4$) is given through the calculation code based on the Gibbs free energy minimization. Finally, the perspectives of this preliminary study will be given in order to propose this new level one test case.

## 2. EXPERIMENTAL SET-UP

The ICP-T64 torch located at the L.A.E.P.T. (Thermal Plasmas and Electrical Arc Laboratory) in Clermont-Ferrand, France, is a classical ICP torch able to work with different kinds of plasma gas (air, argon, $CO_2$, $N_2$ and gas mixtures). A seven-turn induction coil, cooled by air, is used to ignite (with a metallic wire) and sustain the $N_2$-$CH_4$ plasma.

The main features of the experimental set-up and operating conditions are reported in Fig. 1 and Tab. 1, respectively. The plasma is generated through the induction coil by a radio frequency (RF) of 64 MHz delivering a power up to 3 kW. The plasma is confined within a 28-mm quartz tube. The plasma gas is injected at a fixed rate of 0.2 $g.s^{-1}$.

The optical set-up, placed at 34.3 cm from the plasma axis, leads to a spatial resolution of 1mm. Spectral lines intensities are measured with a 0.50 m focal length Czerny-Turner monochromator connected to a CCD detector (1242×1152 pixels, each pixel having a width of 22.5 micrometers). An 1200 grooves/mm grating is used. Its apparatus function $\Delta\lambda_{app}$, assimilated to a Gaussian profile with a Full Width at Half Maximum (FWHM), is calculated from the relation (Eq.1) :

$$\Delta\lambda_{app} = FWHM\ f_p D^{-1} \qquad (1)$$

where FWHM is expressed in pixels, $D^{-1}$ represents the gratings dispersion (1.56 nm/mm) and $f_p$ defines the pixel dimension (22.5 µm). The calculated value of $\Delta\lambda_{app}$ gives 0.14 nm for an entrance slit equal to 100 µm.

**Table 1.** Main features of the experimental set-up.

*Inductively coupled plasma*
Manufacturer/type: Défi Systèmes/ICP T64
Power supply: 64 MHz, 3 kW
Tuning: Automatic adaptation
Inductor: Seven-turn air-cooled coil
Plasma gas flow rate: 0.2 g.s $^{-1}$
Operating pressure: atmospheric pressure
Torch: 28 mm internal diameter quartz tube

**Optical set-up**
Spectrometer: Chromex 500 IS
500 mm focal length
Czerny-turner mounting
Entrance slit: e = 100 µm
Gratings: 1200 grooves.mm$^{-1}$
Detector: CCD EEV 1152×1242 pixels
Spatial resolution: 0.5 mm
Apparatus function: 0.14 nm

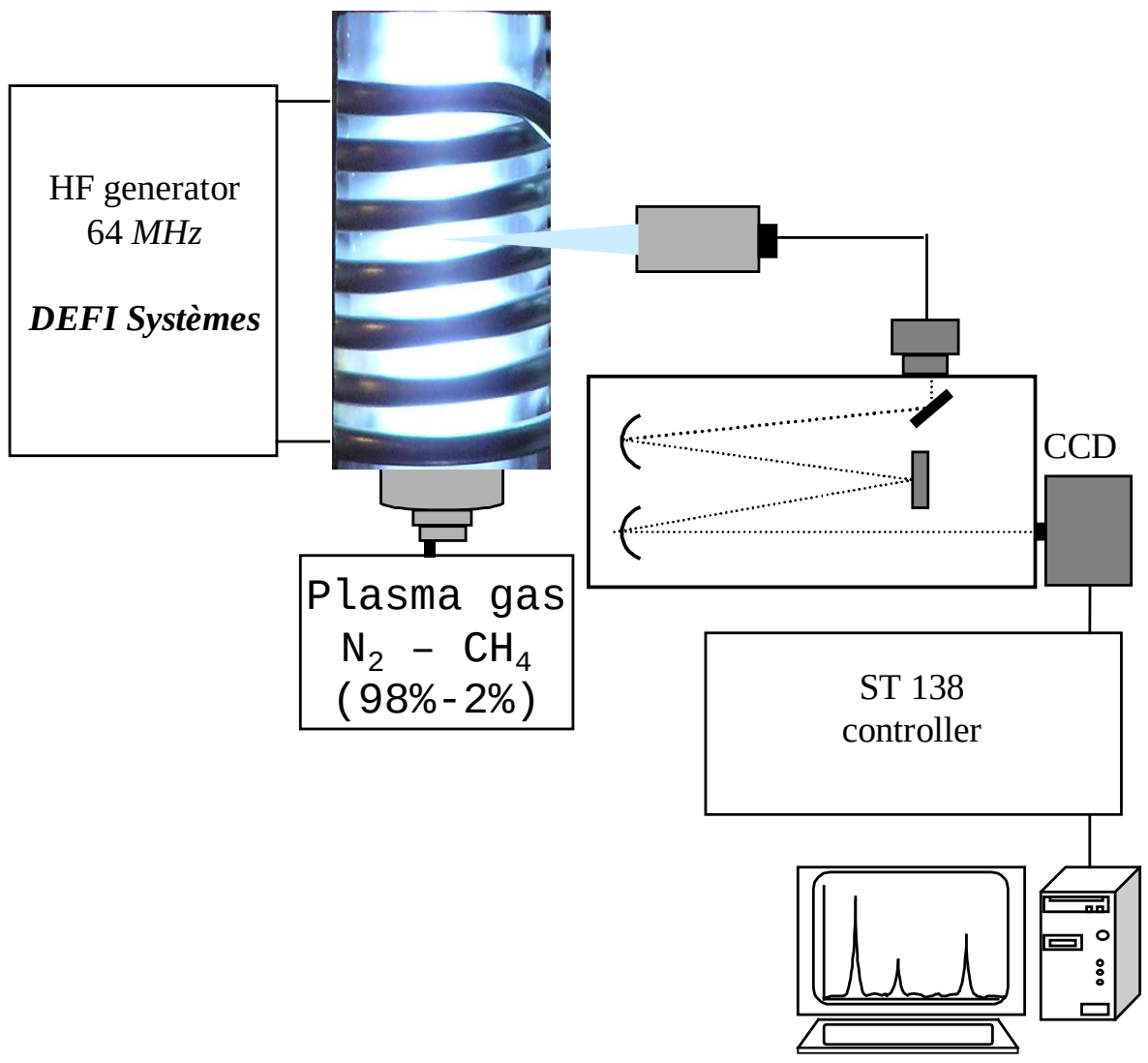

**Figure 1.** Plasma torch and detection experimental set-up.

The measured emission spectra must be corrected from the spectral response of the optical device, which include lens, monochromator and CCD detector. So, a calibration in intensity, between 300 and 800 nm, is necessary to take into account all these effects. A tungsten lamp (PHILIPS WI14) is used to cover the range [300 – 800] nm. The calibration procedure is realized in the same conditions that ones considered for the experimental spectral acquisitions.

## 3. EXPERIMENTAL SPECTRUM

Optical emission spectroscopy is the simplest non-intrusive method for plasma diagnostics. The analytical zone is limited between the fourth and the fifth induction coil where the luminous response is the greatest. Each spectrum corresponds to one acquisition of one second and the electronic noise is automatically subtracted.

One particularity of the Titan atmosphere plasma created in the ICP-T64 torch is the difficulty to have a sustainable plasma. At first, the plasma was ignited with pure nitrogen, and when the methane was injected the plasma generally was cut. It has been observed also that there was a loss of injected power, about 1 kW, when the methane was added. This effect is mainly due to the automatic adaptation of impedance but it is worth noting that there is no problem of extinction when the plasma changes to Titan atmosphere to pure nitrogen. The studies have to be improved to understand this unusual behavior of this kind of plasma.

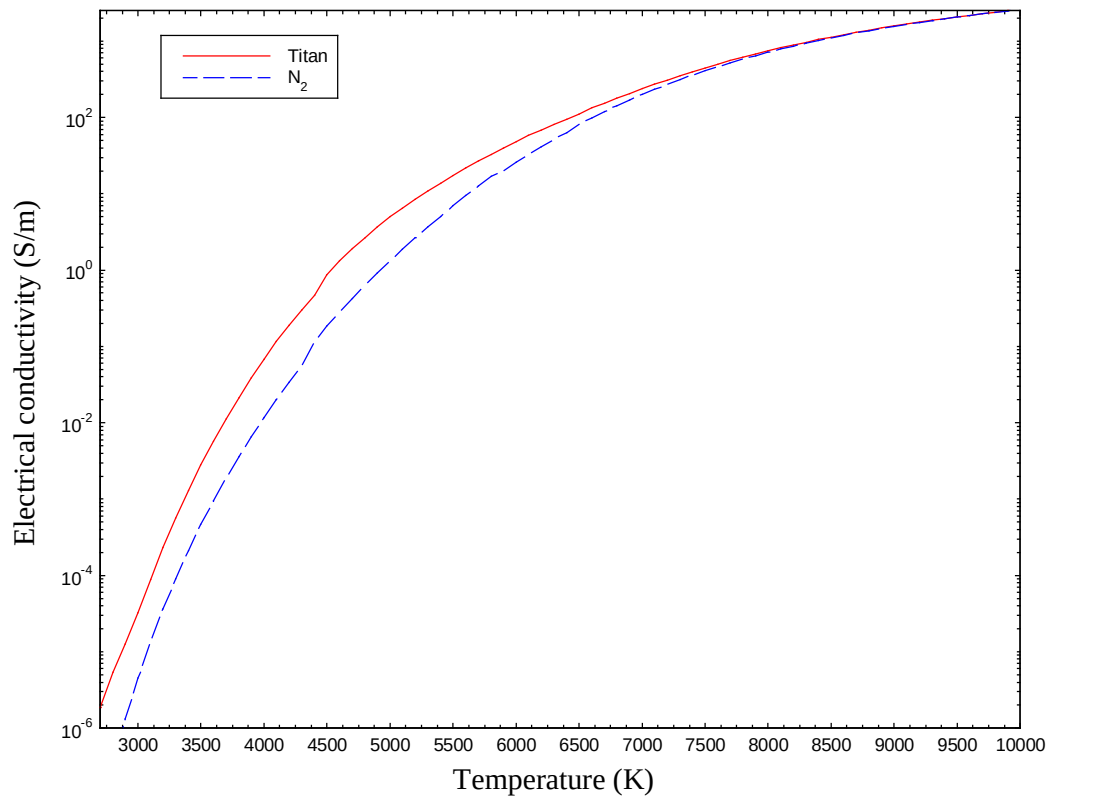

**Figure 2.** Evolution of the electrical conductivity as a function of temperature.

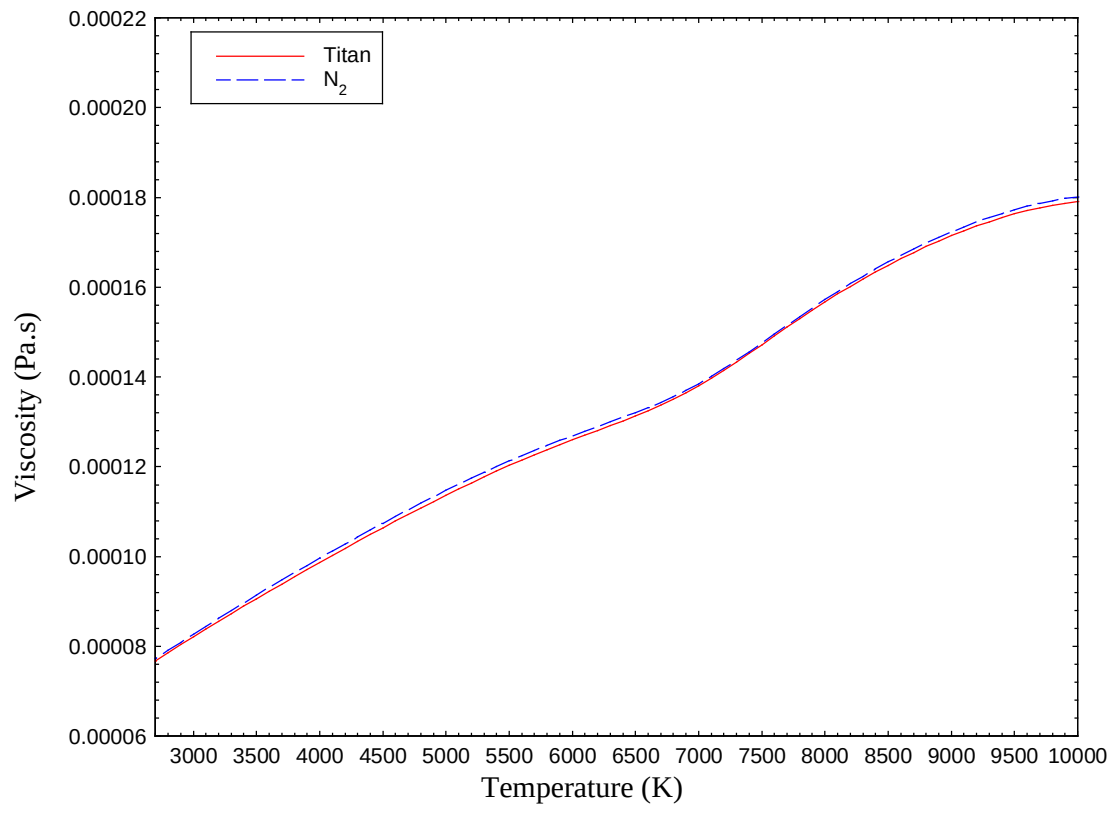

**Figure 3.** Evolution of the viscosity as a function of temperature.

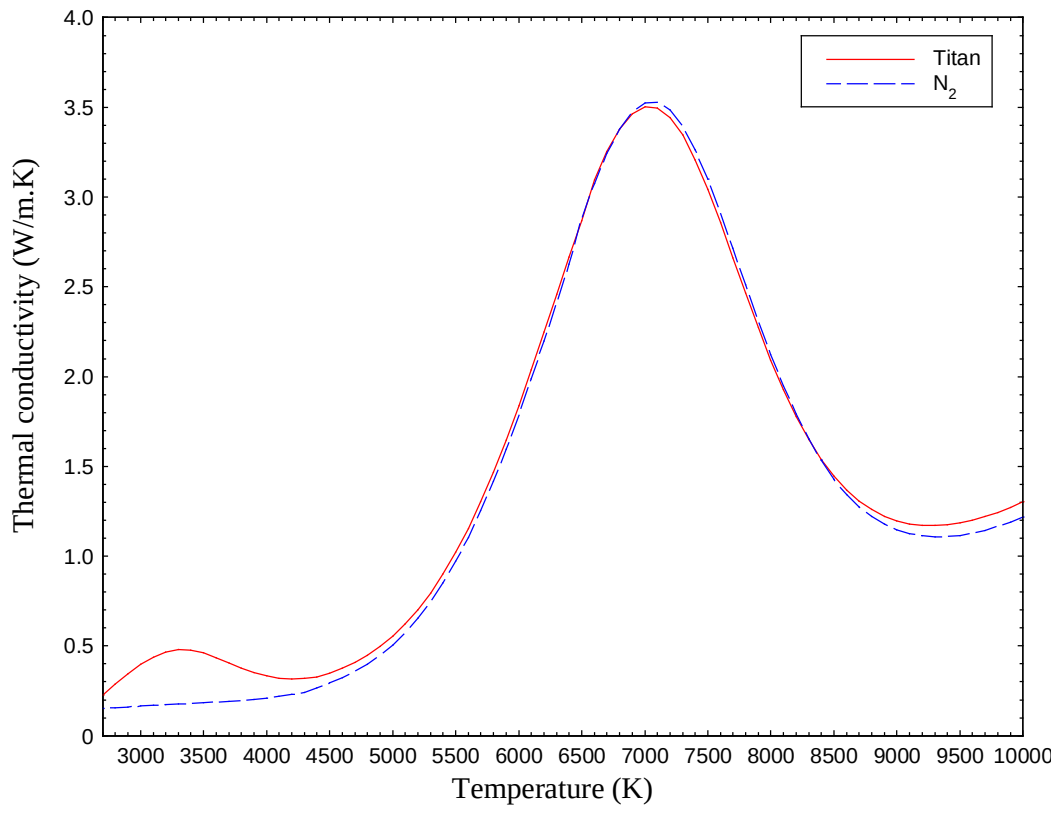

**Figure 4.** Evolution of the thermal conductivity as a function of temperature.

First explanations can be given by considering the different physical properties of the two plasmas (pure $N_2$ and Titan composition).

Fig 2-3-4 report the differences observed for the electrical conductivity, the viscosity and the thermal conductivity of the two plasmas between 2700K and 10000K.

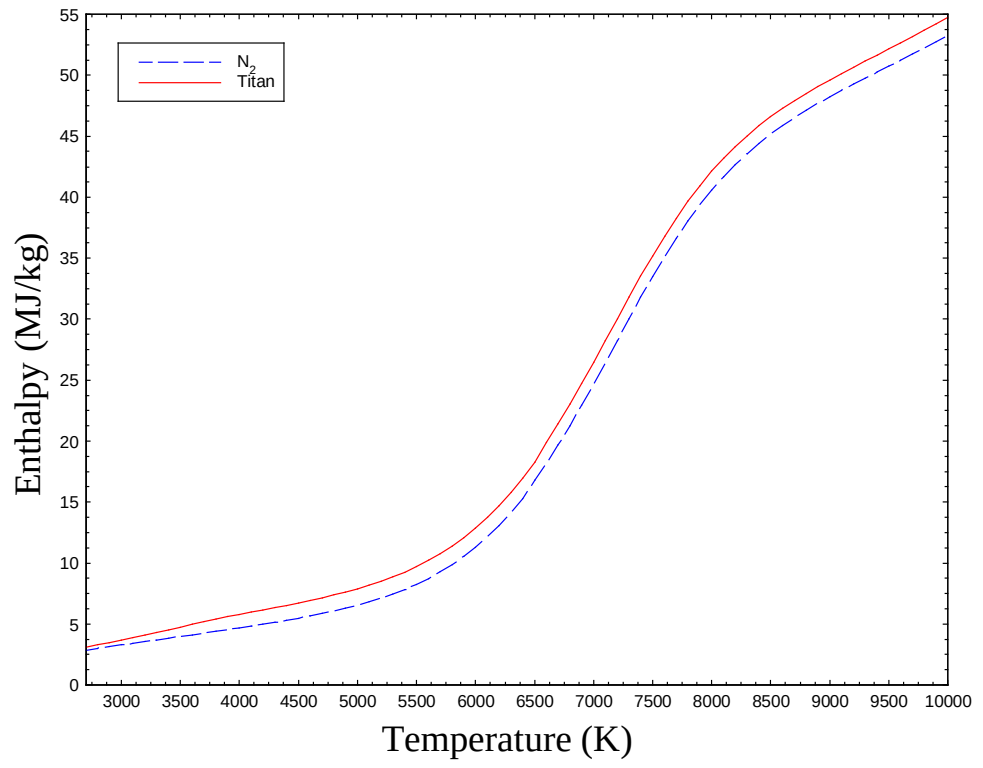

**Figure 5.** Evolution of the specific enthalpy as a function of temperature.

As the viscosity doesn't seem to be an influent parameter, the two others may explain the difficulty to have a viable plasma when methane is added. Concerning the enthalpy (Fig. 5), no significant differences appear and for instant, no clear conclusion can be brought on the specific behaviour of the plasma when methane is added. More thorough studies must be realized.

Fig. 6 reports a rough spectrum recording from the plasma formed with a Titan-like atmosphere in the range [300 – 800] nm. The CN molecular emission is predominant and the $C_2$ emission is also present. No atomic line is observed in the spectral domain.

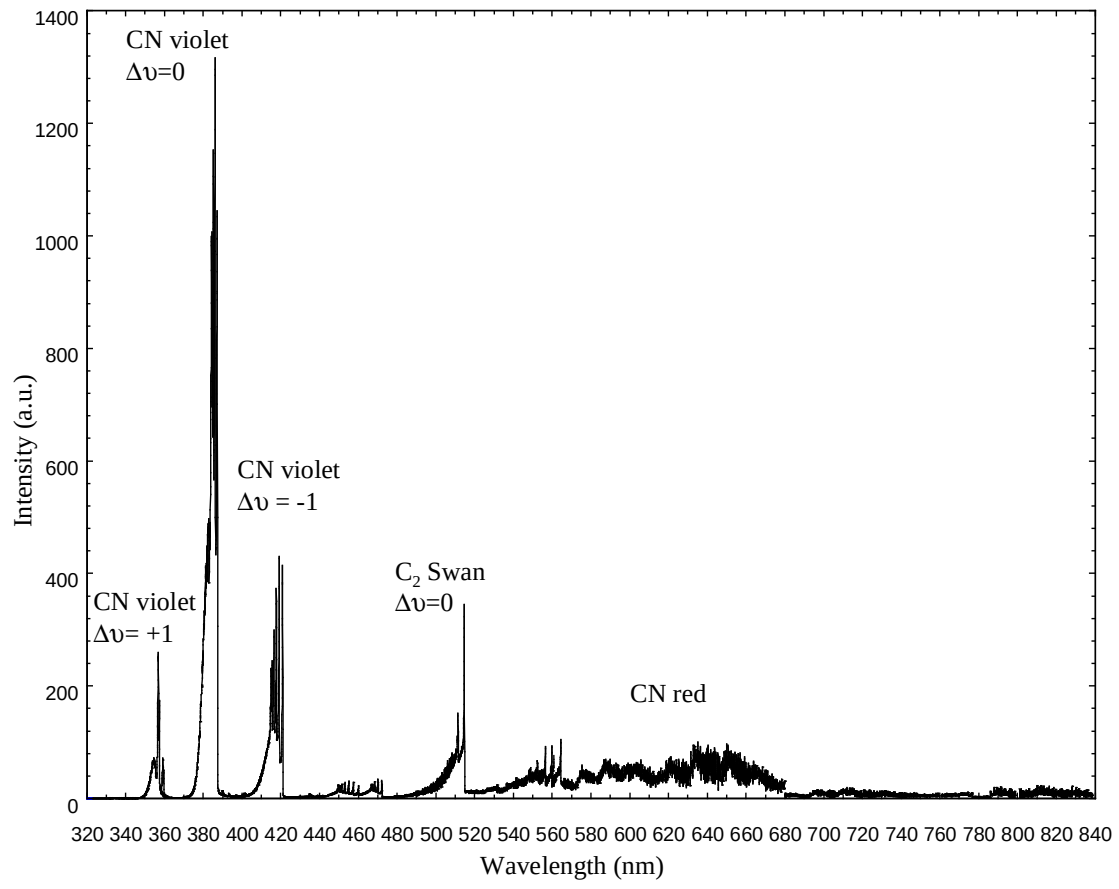

**Figure 6.** Experimental emission spectra for a Titan atmosphere plasma at atmospheric pressure.

Fig. 7 represents different intensity radial evolutions of CN and $C_2$ peaks respectively and Fig. 8 shows the result of the application of Abel inversion. It must be noted that each intensity radial evolution of wavelength is interpolated by cubic splines and fitted by a polynomial of degree 9 before applying the Abel inversion.

It can be noted that the application of Abel inversion is easily practicable for the CN system whereas for the $C_2$ one, very big uncertainties are present. The presence of a plateau of intensity in the axis of the plasma leads to difficulties in order to apply the Abel inversion.

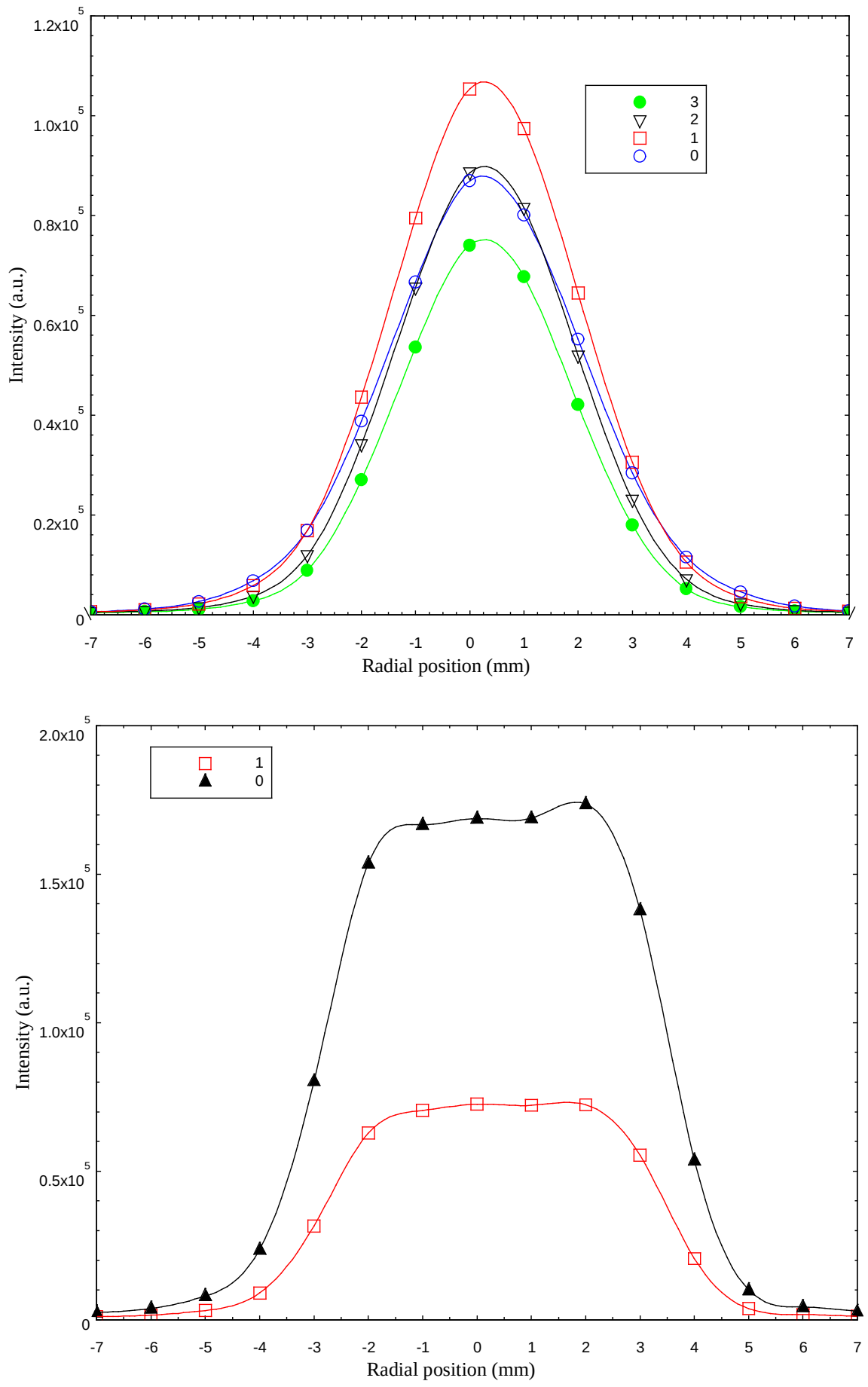

**Figure 7.** Emission profile measured over the plasma diameter for the first four band heads of the CN violet system (~388 nm) and the first two band heads of the $C_2$ swan system (~510 nm).

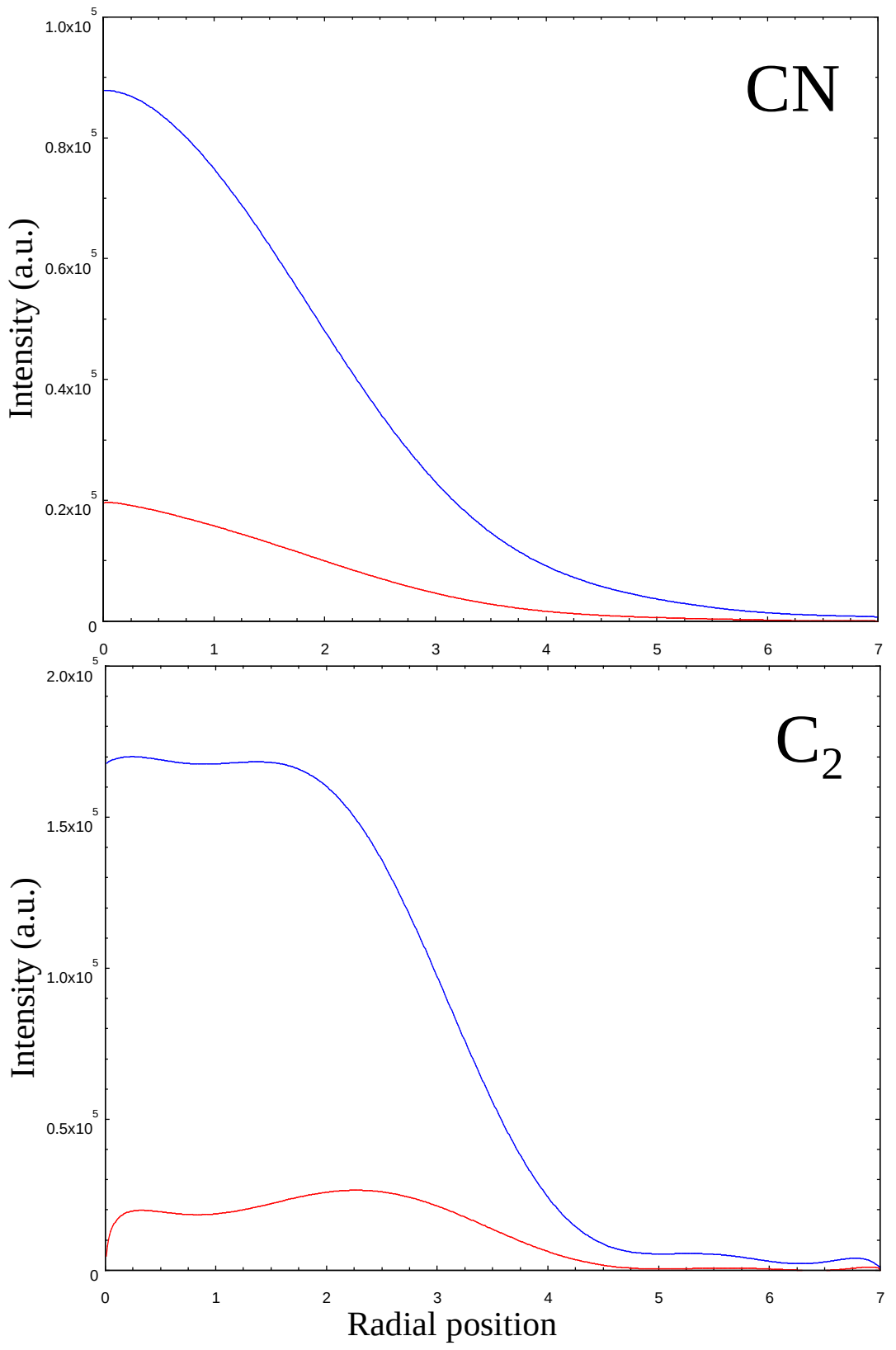

**Figure 8.** Application of the Abel inversion for the first band head of the CN violet system (~388 nm) and the $C_2$ swan system (~510 nm).

In order to estimate the value of temperature, the spectra of the $C_2$ Swan system and the CN violet system has been considered and the SPARTAN code [3]used.

Fig. 9 reports the best agreement obtained by the comparison between the experimental spectrum at the centre of the plasma for the $C_2$ Swan system and the synthetic one. As it is considered that the plasma is at or close to thermal and chemical equilibrium, two temperatures have been estimated in order to have the best agreement, the rotational temperature ($T_r$) and the vibrational one ($T_v$). The obtained values are $T_r$=3200K and $T_v$=3700K.

Fig.10 reports the results concerning the CN violet system. When the experimental spectrum is observed, one can conclude that there is a strong absorption of the lines, especially for the first band head (0-0). So, in order to take into account this absorption, the following calculations are made :

- Estimation of $T_v$ and $T_r$ followed by the calculation of emission $\varepsilon_n$ and absorption $\alpha_n$ coefficients using the line-by-line code SPARTAN;
- Calculation of the slab emitted intensity (Eq. 2)

$$I = \varepsilon_n / \alpha_n \times \left(1 - \exp(\alpha_n \times l)\right) \qquad (2)$$

- Convolution with a gaussian apparatus function simulating the slit.

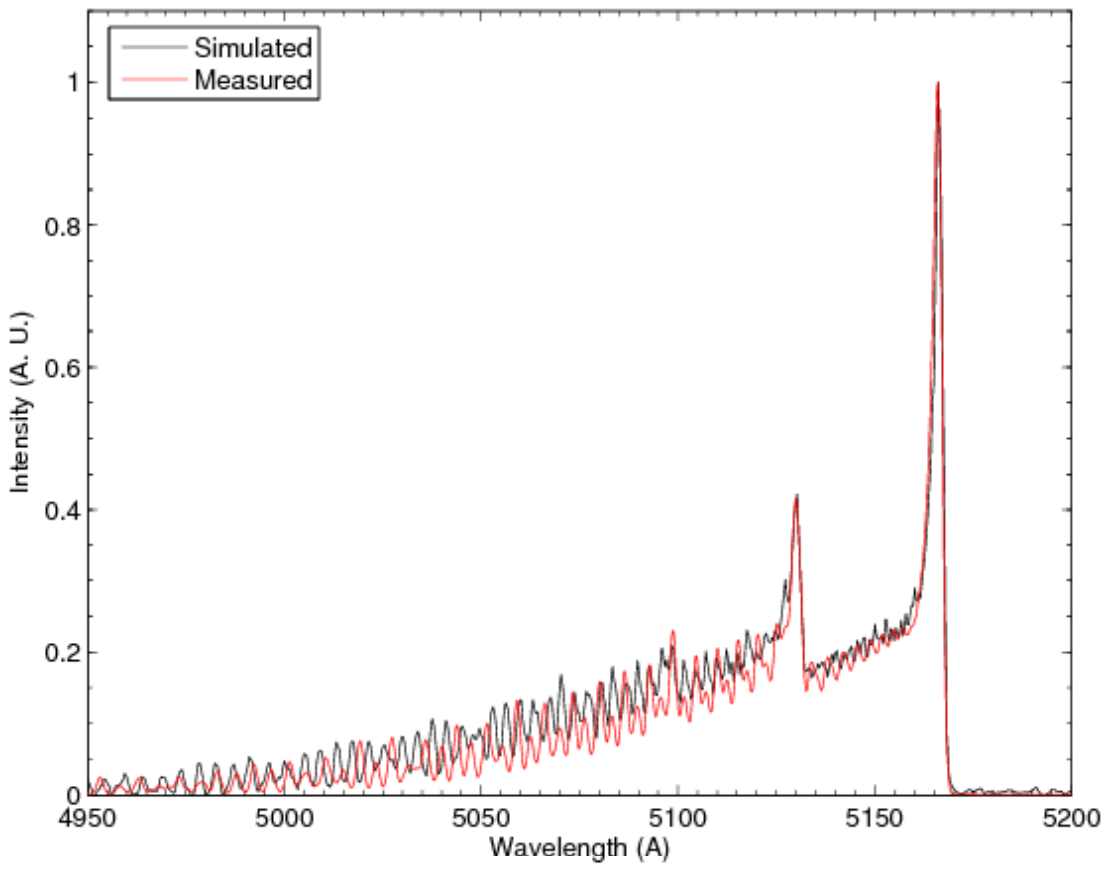

**Figure 9.** Simulation of the radiative emission of the $C_2$ Swan system over the ICP torch axis.

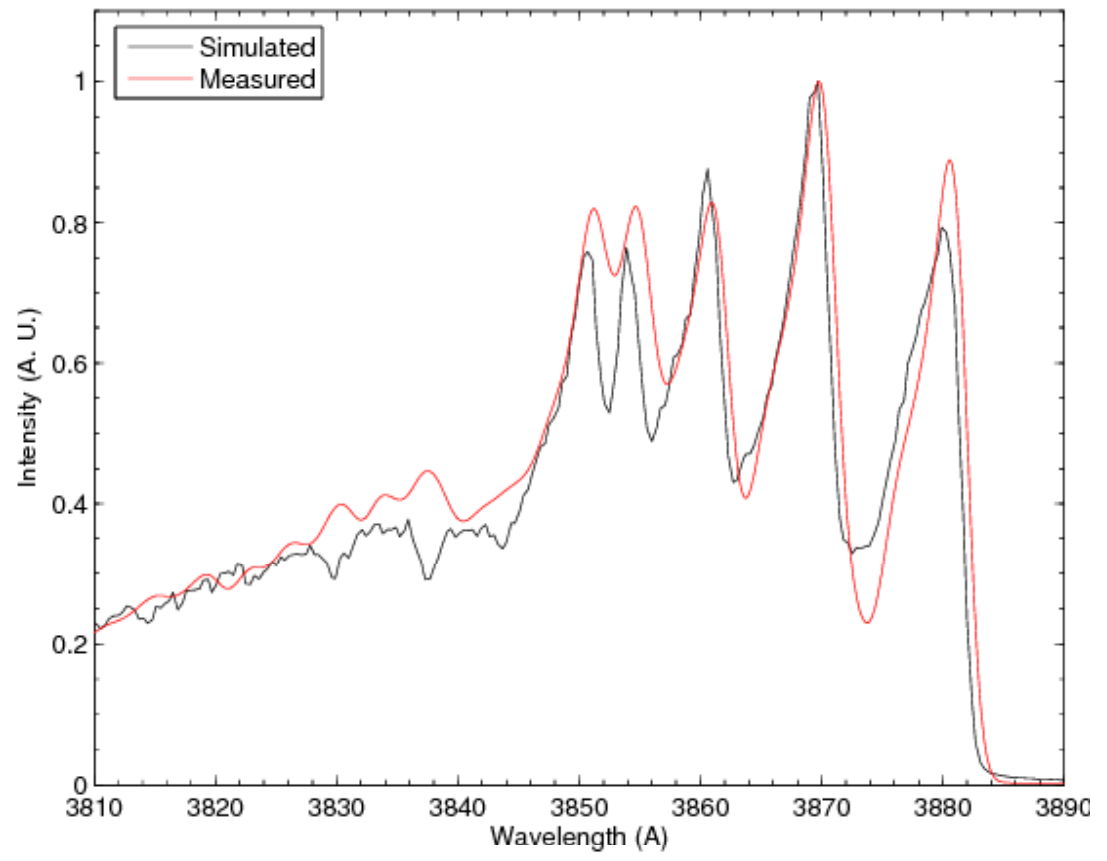

**Figure 10.** Simulation of the radiative emission of the $C_N$ violet system over the ICP torch axis for the $\Delta\upsilon$=0 transitions (best agreement obtained).

The agreement between simulation and experience is not ideal, as it can be seen in Fig. 10. The best gives $T_r$=3200K and $T_v$=3700K with a somehow large uncertainty of 500K. However, this first estimation of temperature, based on the molecular spectra, leads to acceptable results and is a good starting point for future investigations. It is worthy to note that owing to the large uncertainty on the temperatures, it is possible that a thermal equilibrium ($T_r$=$T_v$) might be reached in the plasma.

## 4. DISCUSSION ABOUT THE NUCLEATION PHENOMENON

When the ICP torch works with a Titan like atmosphere, it is observed inside the quartz tube the formation and deposition of "black suit" on the injector (Fig.11) and also on the walls of the quartz tube. It can be noted that the rotation of the deposit follows the movement of vortex of the gas flow.

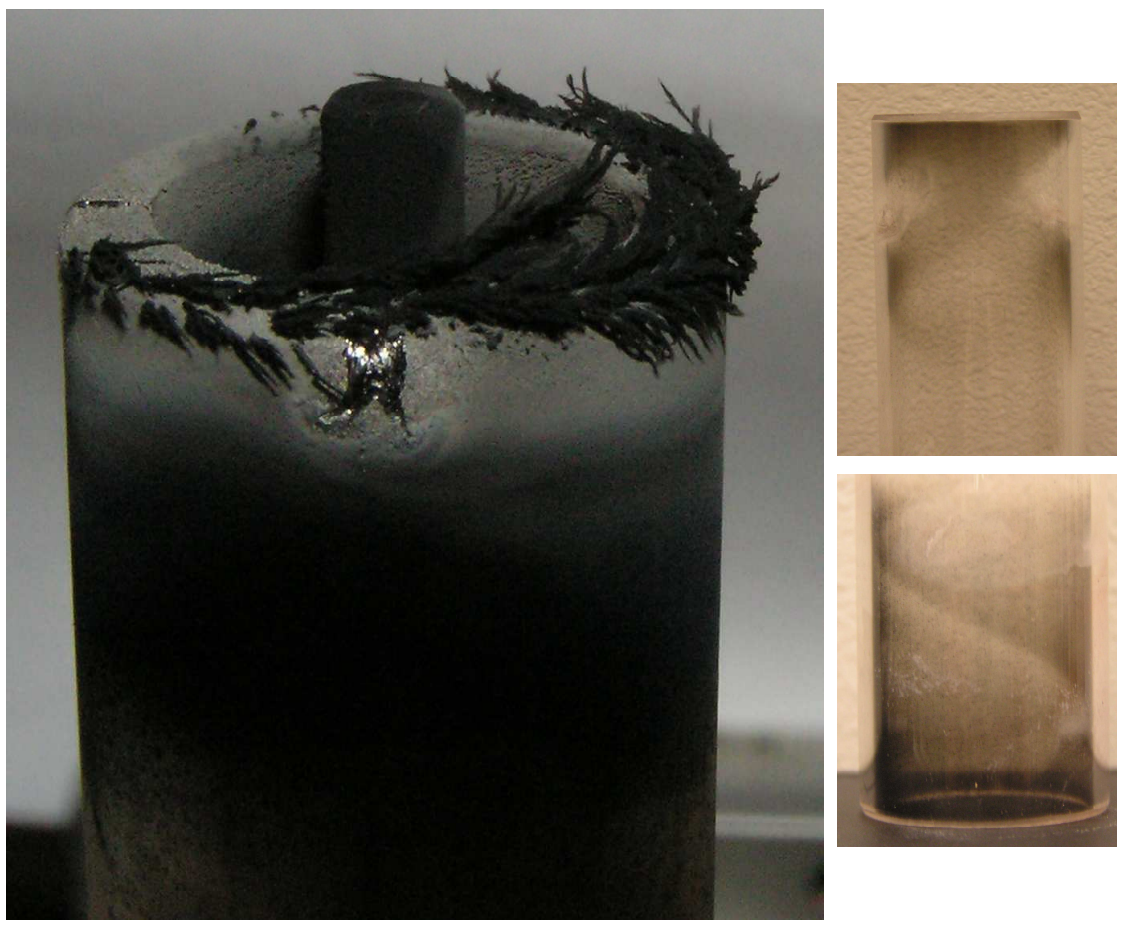

**Figure 11.** Photographs of the suit deposited on the injector (left) and on the walls of the quartz tube (right).

In order to determine what kind of particules was formed in the Titan plasma, an analysis of the suit has been made by using Scanning Electron Microscopy (SEM) with field effect (CASIMIR society). Fig.12 shows a zoom of the particular aspect of the suit.

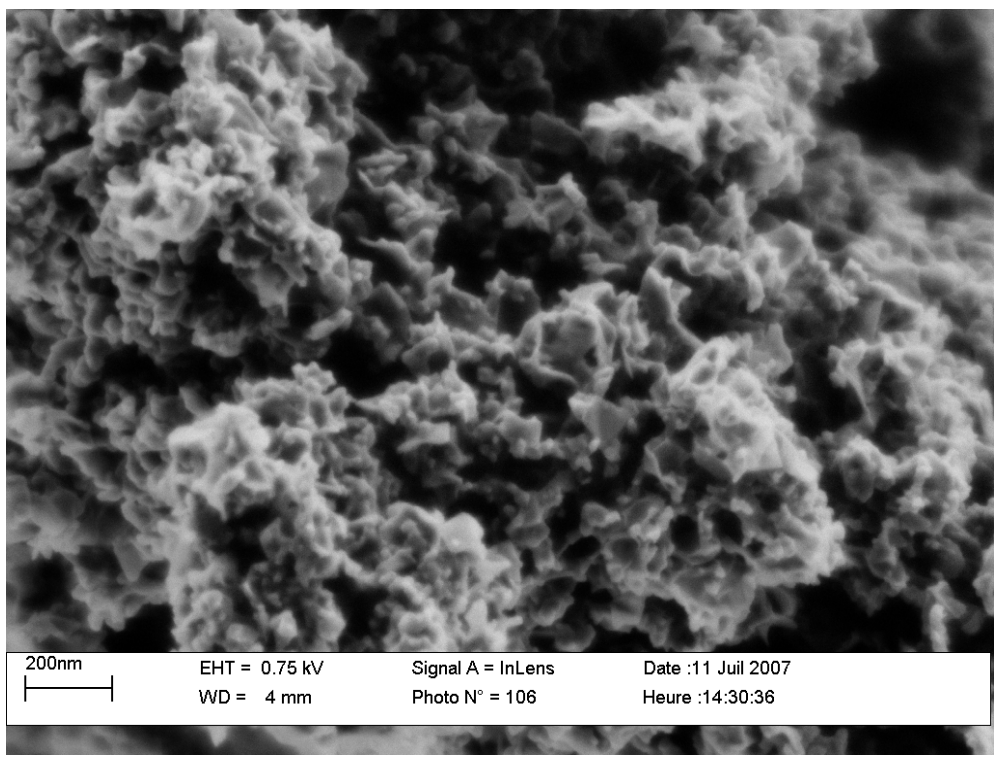

**Figure 12.** Photographs of the suit analysed by SEM with a scale of 200nm.

If the zoom is further increased, spherical particles can be isolated with a diameter of around

35 nanometers. The studies which have been realized on carbon nano-particles synthesis [4] shows similar images to that of Fig.12. This particular aspect which represents the type of the carbon compounds nano-structured is called "ruffled paper structure". An X-ray analysis must be done to determine the true nature of the particules and to verify if it is pure carbon.

From the calculation of the plasma composition based on the Gibbs free energy minimization, one can give a first explanation about the presence of carbon in the plasma torch (Fig. 13).

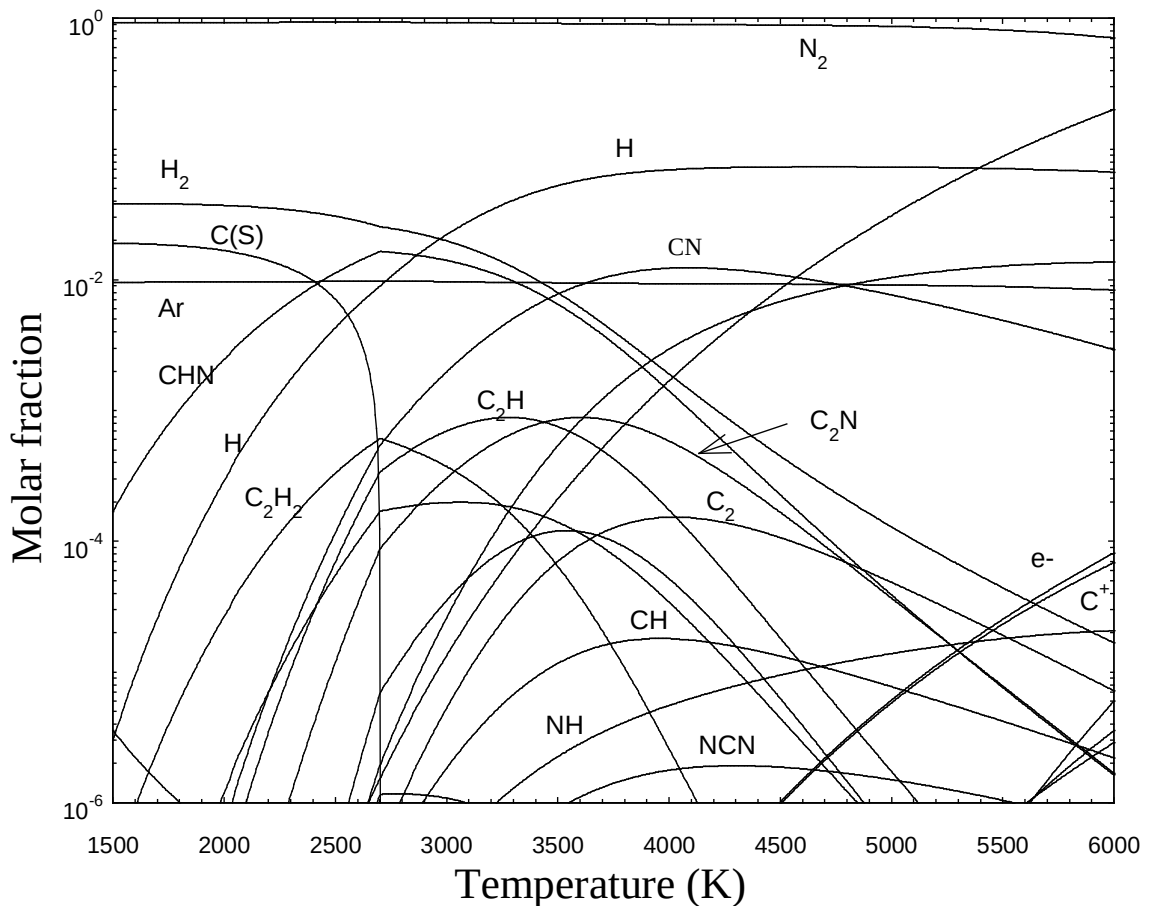

**Figure 13.** Chemical composition (molar fraction) of a $N_2$(98%)-$CH_4$(2%) plasma at equilibrium at atmospheric pressure as a function of temperature.

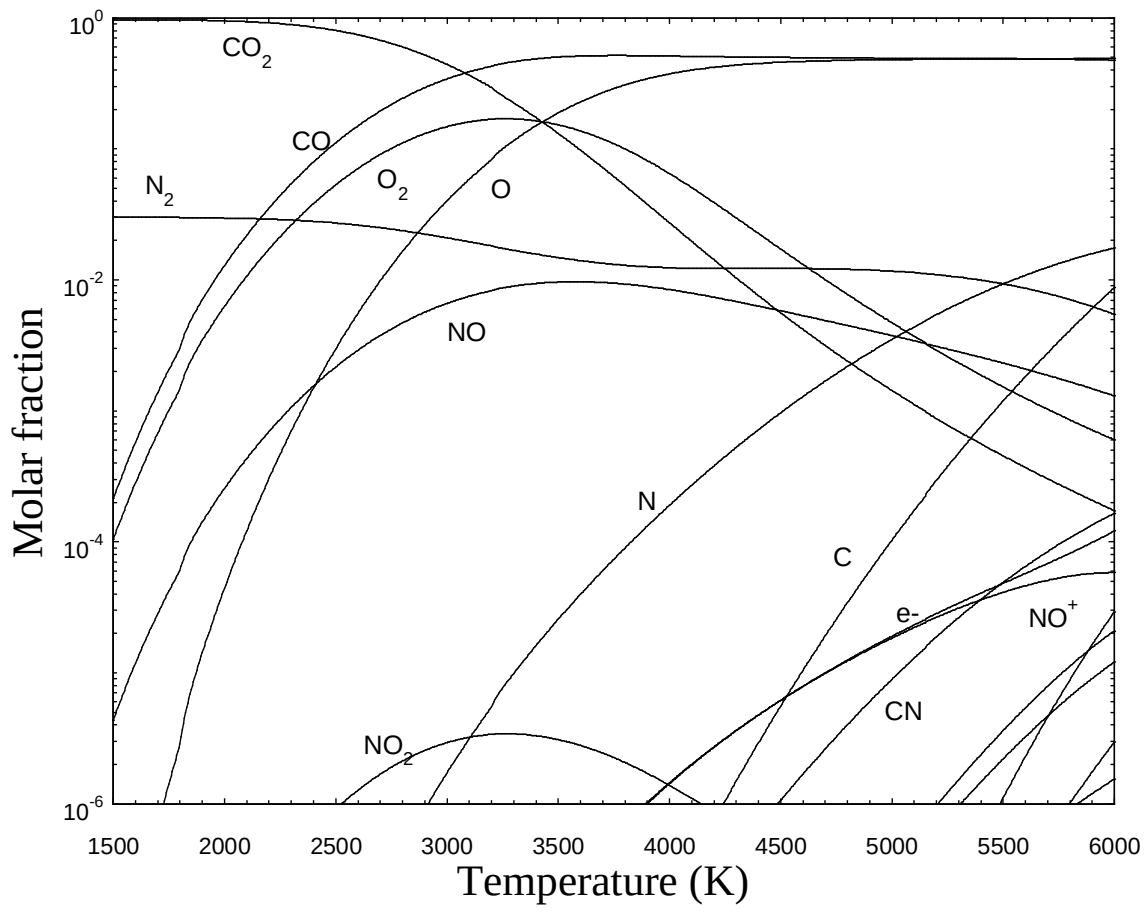

**Figure 14.** Chemical composition (molar fraction) of a $CO_2$(97%)-$N_2$(3%) plasma at equilibrium at atmospheric pressure as a function of temperature.

As it can be observed in the Fig.13, there is the presence of a graphite phase [C(S)] until a temperature of 2700K. So, it confirms the formation of C on the walls of the quartz tube and on the injector which is placed just under the first inductor coil. By comparison with the Mars plasma which has been studied in a previous work and where no formation of carbon has occurred, the Fig. 14 shows also the fact that there is no formation of carbon specie. This absence is mainly due to the presence of $O_2$.

## 4. CONCLUSION

The first investigations on the study of a Titan like atmosphere plasma formed with an inductively coupled plasma torch lead to interesting observations and results, especially the formation of carbon species. It was also shown that the addition of methane strongly disturbed the plasma and explanations about this change of behaviour have to be carried out.

Temperatures have been estimated but only at the centre of the plasma. Further works have to be done in order to determine the radial evolution of temperature and if a weak departure from equilibrium exists in this kind of plasma. For this, the addition of a very small percentage of argon in the plasma would be interesting as we could have atomic lines to determine the atomic excitation temperature.

A new optical diagnostic has to be added to the experimental set-up in order to determine the electronic density (Laser Interferometry). Moreover, it would be of interest to realize kinetic calculations to confirm that the chemical equilibrium is reached inside the inductor where the optical measurements are recorded (as it has been done for the study of the Mars plasma).

The aim of the future investigations is so to propose a complete level-one test case for the validation of radiative codes simulating equilibrium radiation from a Titan type plasma in equilibrium conditions in the near UV to near IR range.

ACKNOWLEDGMENTS

This research work has been carried out thanks to the financial support of the RayHen ANR and the support of the European Space Agency ESA in the scope of the AURORA program.